\documentclass[hyper]{JHEP3}
\pdfoutput=1
\usepackage{amsmath}
\usepackage{epsfig}
\usepackage[all]{xy}
\usepackage{feynmf}
\usepackage{slashed}

\newcommand{\CA}{\mathcal{A}}

\newcommand{\CF}{\mathcal{F}}

\newcommand{\CK}{\mathcal{K}}

\newcommand{\CN}{\mathcal{N}}

\newcommand{\CD}{\mathcal{D}}

\newcommand{\CV}{\mathcal{V}}

\newcommand{\SU}{\mathrm{SU}}

\newcommand{\USp}{\mathrm{USp}}
\newcommand{\U}{\mathrm{U}}

\newcommand{\half}{\frac{1}{2}}
\newcommand{\ndt}{\noindent}

\def\p{\partial}

\def\bea{\begin{eqnarray}}
\def\eea{\end{eqnarray}}
\def\be{\begin{equation}}
\def\ee{\end{equation}}
\def\ba{\begin{align}}
\def\ea{\end{align}}
\def\bse{\begin{subequations}}
\def\ese{\end{subequations}}
\newcommand{\ben}{\begin{eqnarray}}
\newcommand{\een}{\end{eqnarray}}

\newcommand{\bem}{\begin{pmatrix}}
\newcommand{\eem}{\end{pmatrix}}

\def\={\;  = \;}
\def\+{\, + \,}

\def\wt{\widetilde}
\def\wh{\widehat}
\def\bar{\overline}

\def\rt2{\sqrt{2}}

\def\s{\sigma}
\def\g{\gamma}

\def\a{\alpha}
\def\b{\beta}
\def\d{\delta}

\def\eps{\epsilon}

\def\O{{\Omega}}
\def\G{\Gamma}

\def\ve{\varepsilon}
\def\vp{\varphi}

\def\nv{n_{\rm v}}

\title{All solutions of the localization equations for $\CN=2$ quantum black hole entropy}

\preprint{NIKHEF2012-009}

\author{Rajesh Kumar Gupta and Sameer Murthy\\
NIKHEF theory group, Science Park 105, \\
1098 XG Amsterdam, The Netherlands \\

{\tt \email{rgupta, smurthy @nikhef.nl}, \email{murthy.sameer@gmail.com}} }

\abstract{ 
We find the most general bosonic solution to the localization equations describing the 
contributions to the quantum entropy of supersymmetric black holes in four-dimensional
$\CN=2$ supergravity coupled to $n_{\rm v}$ vector multiplets. 
This requires the analysis of the BPS equations of the corresponding off-shell supergravity 
(including fluctuations of the auxiliary fields) with $AdS_{2} \times S^{2}$ attractor boundary conditions.
Our work completes and extends the results of arXiv:1012.0265 that were 
obtained for the vector multiplet sector, to include the fluctuations of 
all the fields of the off-shell supergravity.  
We find that, when the auxiliary~$SU(2)$ gauge field strength vanishes, the most general 
supersymmetric configuration preserving four supercharges is labelled by $n_{\rm v}+1$ 
real parameters corresponding to the excitations of the conformal mode of the graviton and 
the scalars of the $n_{\rm v}$ vector multiplets. In the general case, the localization manifold 
is labelled by an additional $SU(2)$ triplet of one-forms and a scalar function.
}


\begin{document}

\section{Introduction and summary}

Recently, there has been quite some progress in computing the quantum entropy of black holes 
\cite{Banerjee:2010qc, Banerjee:2011jp, Sen:2011ba, Sen:2011cj, Dabholkar:2010uh, Dabholkar:2011ec}, 
extending precision calculations of classical black hole entropy (see \cite{Mandal:2010cj} for a 
recent review) to include quantum effects. 
This quantity is computed in the gravitational theory and is to be considered as the 
quantum generalization of the classical Bekenstein-Hawking-Wald entropy 
\cite{Bekenstein:1973ur, Hawking:1974sw, Wald:1993nt, Iyer:1994ys, Jacobson:1994qe} of black holes. 
In the context of supersymmetric black holes in theories of supergravity, 
an \emph{exact} computation of the quantum entropy was performed in 
\cite{Dabholkar:2010uh, Dabholkar:2011ec}, 
effectively summing over all the perturbative quantum fluctuations of the theory at one shot. 
The computation of the exact quantum black hole entropy allows us to compare it 
with the exact statistical entropy of an ensemble of states with the same charges, 
the latter being the logarithm of an integer (the degeneracy of states). 
The equality of the two notions of entropy is a universal expectation in any purported 
consistent quantum theory of gravity such as string theory, and thus the above comparison 
leads to a very stringent test of the theory.

The computation of \cite{Dabholkar:2010uh} uses the definition of the exact quantum entropy of extremal black holes 
that has been proposed by Sen in \cite{Sen:2008yk, Sen:2008vm}.  
The quantum entropy of a black hole with charge vector $(q,p)$ is defined as the logarithm of the quantum 
expectation value $W(q,p)$ of a Wilson line inserted on the boundary of the $AdS_{2}$ space that 
appears as a factor in the near-horizon region of the extremal black hole.
The quantum expectation value is defined as a Euclidean functional integral over all
the quantum fields in the theory on $AdS_{2}$, and like any other 
functional integral, it is an enormously complicated object.

The idea of \cite{Dabholkar:2010uh} was to use localization techniques 
\cite{Witten:1988ze, Witten:1991zz, Witten:1991mk, Schwarz:1995dg, Zaboronsky:1996qn}  
to simplify this functional integral and reduce it to a finite number of ordinary integrals. This is 
done in the context of four-dimensional $\CN=2$ supergravity coupled to an arbitrary number 
$\nv$ of vector multiplets. In the classical theory, the supersymmetric black hole 
configuration is fixed by the attractor mechanism -- the geometry is $AdS_{2} \times S^{2}$,  
the vector fields have constant field strengths, and the scalar fields have 
constant values that are determined by the charges. 
In the quantum functional integral, the fields are held fixed to their attractor values at the boundary 
of $AdS_{2}$, and they are allowed to fluctuate in the interior.

The idea of localization is that if the quantum theory admits a conserved supercharge~$Q$, 
then the functional integral collapses or \emph{localizes} to an integral over the space of 
solutions of the equation $Q\psi=0$, where $\psi$ denotes the fermions of the theory. 
This \emph{localization manifold} is often enormously smaller than the original configuration 
space of the fields of the theory. By applying localization to the $\CN=2$ supergravity
theory, an exact formula for the quantum entropy was derived in  \cite{Dabholkar:2010uh}. This formula 
was inspired by, and is similar in spirit to the formula of \cite{Ooguri:2004zv}, but there are differences 
that were spelt out  in  \cite{Dabholkar:2010uh}. 
A concrete application of the exact quantum entropy formula of \cite{Dabholkar:2010uh} 
to black holes in $\CN=8$ string theory yielded successful results -- 
the black hole degeneracy that was computed in~\cite{Dabholkar:2011ec} by this method was 
exponentially close to the exact microscopic result~\cite{Maldacena:1999bp}. 

At the level of a derivation, however, the analysis of \cite{Dabholkar:2010uh} included 
some assumptions that have not been proven (even at a physicist's level) so far. 
One of the assumptions concerns the structure of the  localization manifold, which 
involves finding all solutions of the localization equations $Q\psi=0$.  
In \cite{Dabholkar:2010uh}, a restricted ansatz was used for the vector multiplet sector, 
within which all the solutions were found. Furthermore, in the gravity multiplet 
sector, it was assumed that the only bosonic configurations that contribute  
are the fluctuations of the conformal mode of the graviton. 
In this paper, we perform a complete analysis of the localization equations for the~$\CN=2$
supergravity coupled to an arbitrary number of vector multiplets.
As we shall explain below, the answer obtained in~\cite{Dabholkar:2010uh} is indeed the 
complete localization manifold when a certain auxiliary gauge field is set to zero, and otherwise, 
the manifold is larger. We comment on the implications of this result in the following.  

In order to apply localization to problems of this sort, one needs a formalism in which the 
supersymmetry algebra closes off-shell. As in \cite{Dabholkar:2010uh}, we shall use the formalism of conformal 
supergravity which allows for off-shell closure of the algebra on all the fields of the 
gravitational and vector multiplets. 
In this formalism, the supersymmetry transformations are independent of the 
action, {\it i.e.}~they do not depend on the prepotential of the $\CN=2$ theory nor on the 
higher derivative terms in the Lagrangian. In this sense, the solutions to the BPS equations 
in the gravity and vector multiplet sector that we will find are universal, and 
given a particular action, one can evaluate and integrate it on this space of solutions.

Before we discuss our technique and present our results, 
we briefly comment on two important issues concerning localization in supergravity. 
The first issue regards the configuration space in which we look for solutions of the 
equation $Q\psi=0$. 
Our working assumption is that string theory provides a consistent ultraviolet cutoff 
to the functional integral in gravity. We should, therefore, include all configurations 
which are allowed in classical string theory. Since such a classification remains 
to be done, we shall look for smooth configurations in this paper, and leave ourselves open to the 
possibility that there may be classically singular configurations that are consistent in string 
theory\footnote{In many situations \cite{Banerjee:2008ky, Murthy:2009dq}, it indeed 
seems to be the case that orbifolds should be included in the functional integral to 
recover the exponentially suppressed contribution to the entropy.}.

A second important issue is that of background independence\footnote{We thank
Atish Dabholkar, and especially Jo\~ao Gomes for emphasizing this point to us.}. A key step 
in the mechanism of localization involves the identification of a symmetry of the quantum 
theory. In a theory with a fixed background metric, there is a good definition of symmetries 
(as Killing vectors or spinors of the spacetime). Here, we would like to allow the metric 
to fluctuate, and it is not so clear what the definition of 
Killing vectors and spinors should be. 
Near the boundary of the space, the metric is close to the $AdS_{2} \times S^{2}$ metric, 
and so we can formulate a Killing equation for a small fluctuation around the 
classical metric. However, we cannot do that deep inside the bulk where the metric can fluctuate arbitrarily. 
We take the following attitude inspired by two-dimensional quantum gravity and Liouville theory 
\cite{Ginsparg:1993is}. 
We think of the Killing spinor equation $Q\psi =0$ as being formulated around 
the classical $AdS_{2} \times S^{2}$ metric, and integrate over arbitrary fluctuations of 
the metric. Our assumption (and hope) is that the measure of the functional 
integral is independent of this background choice of metric, or at least lies in the same universality 
class as ``the correct'' quantum theory \`a la Liouville theory.

We now summarize our technique to find all solutions to the localization equations $Q\psi=0$. 
This equation is not easy to solve since we have to solve for $\psi$ as well as the 
bosonic quantities that enter the definition of $Q$. 
The idea of the solution, which goes back to \cite{Tod:1983pm, Biran:1982eg, deWit:1983gs, 
Candelas:1985en, Tod:1995jf}
is to first assume that there is at least one solution 
$\psi$ of the equation. One then forms bilinears $\overline \psi \psi$,  $\overline \psi \g_{\mu}\psi$~etc,
which are all spacetime bosonic tensors. Using the spinor equation $Q\psi=0$, one writes 
first order equations for the bosonic quantities. These quantities can then be identified with 
bosonic objects in spacetime such as the Killing vector (in the case of $\overline \psi \g_{\mu}\psi$). 
Since we can construct as many bosonic degrees of freedom as fermionic ones, we have the 
same content as the original equation, now spread over many first order bosonic equations. 

This equivalent system of equations does not involve spinorial quantities and is therefore  
easier to solve. Having found all the bosonic quantities, we can 
plug them into the original equation to solve for the spinor $\psi$. 
This strategy has been applied with great success to a wide range of on-shell BPS problems 
(see \cite{Gauntlett:2002nw} and follow-ups),
including the four-dimensional $\CN=2$ supergravity coupled to $n_{V}+1$ vector multiplets 
\cite{Meessen:2006tu} which is closely related to our problem. 
Their analysis, however, involves solving for the auxiliary fields using their equations of motion 
(which we are not allowed to do here). Nevertheless, the structural analysis of \cite{Meessen:2006tu}
will be very useful to us, and we shall closely follow their approach. One important feature specific 
to our analysis is that the $AdS_{2}$ boundary conditions are highly constraining, 
and they eliminate many of the fluctuations allowed by the BPS equations.

A related but slightly different method was used in \cite{Dabholkar:2010uh} to solve the 
vector multiplet equations. The idea is to first form the object $(\overline{Q\psi} \, , Q \psi)$, 
which is a positive-definite bosonic quantity.  One writes this as a sum of perfect squares, 
and then sets each perfect square to zero separately. This approach is useful in localization computations since 
one has to make an analytic continuation and the identification as perfect squares suggests the correct 
analytic continuation as in \cite{Pestun:2007rz}. 
The analytic continuation is very important to get our results -- as we shall discuss below, 
a different choice leads to a larger set of solutions. 

Our results are as follows. The~$\CN=2$ off-shell multiplet contains a gauge field for 
the~$SU(2)$ R-symmetry of the theory. When we set its field strength to zero, 
we find that the full set of bosonic solutions to the 
localization equations in $\CN=2$ off-shell supergravity coupled to $\nv$ vector multiplets is labelled 
by $\nv+1$ real parameters. 
These parameters label the size of fluctuations of a certain shape 
(fixed by supersymmetry) of the conformal mode of the metric and of the scalars in the $\nv$ vector 
multiplets, and can be taken to be the 
values of these $\nv+1$ fields at the center of $AdS_{2}$. As mentioned above, this is exactly 
what is needed for a consistency with the microscopic results of string theory. 

The~$SU(2)$ gauge field, on the other hand, is not fixed up to a finite number of parameters 
by the supersymmetry equations. The set of off-shell BPS configurations is parameterized 
by an~$SU(2)$ triplet~$\Phi^{(a)}$ of one-forms, and one scalar function that is the projection 
of the~$SU(2)$ gauge field onto this triplet~$\sum_{a=1}^{3} \CV_{\mu}^{(a)} \Phi^{(a)\mu}$. 
This raises a question about the mechanism of the matching of results between the microscopic 
and gravity computations. In this regard, we note that there is also a~$U(1)$ gauge field in the 
Weyl multiplet which is not fixed by the BPS equations of the Weyl multiplet, but (when~$\CV_{\mu}^{ij}=0$) 
it is localized completely upon coupling to the vector multiplets. This suggests that it may be important to 
consider hypermultiplets in the analysis, which naturally couple to the~$SU(2)$ gauge fields.

A final note is about the possible use of our work in localization problems in other contexts. 
There has been a lot of recent interest in the BPS equations of supergravity motivated 
by applications to supersymmetric field theory \cite{Festuccia:2011ws, Dumitrescu:2012ha, Klare:2012gn}. 
In that case, one takes the Planck mass to infinity to reach the limit of rigid supersymmetry. 
In our case, we do not scale away the fluctuations of any field in the theory, since our motivation 
is to study black holes which involve excitations of the matter fields of the supergravity theory.
However, the supersymmetry variation equations that we solve include those of the 
gravity multiplet, and so our work may have possible applications in studying field theories 
with~$\CN=2$ supersymmetry.

The plan of the paper is as follows. In \S\ref{bpseqns}, we collect the relevant facts about the 
off-shell supergravity formalism that we use. We then discuss the various gauge-fixing procedures 
and write down the explicit supersymmetry equations that we shall solve. 
In \S\ref{Weyl}, we solve the equations in the Weyl multiplet sector,
and in \S\ref{vectors}, we shall review the solution in the vector multiplet sector.
In \S\ref{gaugeanal}, we consider the inclusion of the gauge fields and present the 
modified analysis. 
We present some details of the calculations in the Weyl multiplet and vector multiplets in 
Appendices~\S\ref{smoothV} and \S\ref{vecdetails}.

\section{BPS equations in off-shell supergravity \label{bpseqns}}

We shall study $\CN=2$ supergravity in four dimensions coupled to $\nv$ vector 
mutliplets using the formalism of conformal supergravity \cite{deWit:1979ug, deWit:1984px, deWit:1980tn}, 
(see \cite{Mohaupt:2000mj} for a review). 
This formalism allows us to consider off-shell 
supersymmetry variations which, as explained in the introduction, is of particular interest to us. 
The theory enlarges the Poincar\'e supersymmetry group to the superconformal
group in four dimensions by adding extra fields to the fields of Poincar\'e supergravity.
In the enlarged theory, the supersymmetry algebra closes on all the fields, independent 
of the form of the action. In order to reach the physical theory of interest, we will have 
to gauge fix the extra symmetries, and impose equations of motion on all the auxiliary fields. 
There is a huge literature on this subject, and we shall summarize the points relevant to 
us below, using the above references. 

For our application, we need to do the gauge-fixing of the extra symmetries as usual since we want to consider 
the physical supergravity theory (and not a theory with more symmetries). 
However, we want to keep the fluctuating auxiliary fields in the functional integral, so as to close 
the supersymmetry variations off-shell in order for the localization formalism to be valid. 
Hence, contrary to the usual treatment, we shall not impose the equations of motion for 
the auxiliary fields. In this section, we spell out the details of these steps in 
order to obtain ``gauge-fixed off-shell BPS variations'', and discuss the boundary conditions 
on the various fields entering these equations. In the following sections, we shall find all solutions to these equations.

The basic field content of the theory consists of a Weyl multiplet and $(\nv +1)$ 
vector multiplets. The one extra vector multiplet is needed as a compensating multiplet 
to realize all the symmetries of the off-shell  theory. In addition, one also always needs a second compensating 
multiplet to gauge fix the extra gauge symmetries of the conformal supergravity theory, 
we choose this to be a hypermultiplet as in \cite{deWit:1980tn}. 
Two derivative terms in the Lagrangian are described by a minimal coupling 
of these multiplets, and higher derivative terms can be added to the Lagrangian by introducing 
other chiral multiplets built out of these basic multiplets.

The field content of the \emph{Weyl multiplet} is 
\be\label{Weylfields}
{\bf w} = \left( e_{\mu}^{a}, w_{\mu}^{ab}, \psi_{\mu}^{i}, \phi_{\mu}^{i}, b_{\mu}, f_{\mu}^{a}, A_{\mu}, \CV_{\mu \, j}^{\, i},  T_{ab}^{ij}, \chi^{i}, D \right) \, .
\ee
Here the fields $(e_{\mu}^{a}, w_{\mu}^{ab})$ are the gauge fields for translations and Lorentz transformations;
$\psi_{\mu}^{i}, \phi_{\mu}^{i}$ are the gauge fields for Q-supersymmetries and the  conformal 
S-supersymmetries; 
$(b_{\mu}, f_{\mu}^{a})$ are the gauge fields for dilatations and the special conformal transformations; and 
$(\CV_{\mu \, j}^{\, i}, A_{\mu})$ are the gauge fields for the $\SU(2)$ and $\U(1)$ R-symmetries. 
In the physical theory, $e_{\mu}^{a}$ and $\psi_{\mu}^{i}$ become the vielbein and the gravitini. 
Imposition of the ``conventional constraints'' determines $w_{\mu}^{ab}, \phi_{\mu}^{i},  f_{\mu}^{a}$ in terms of other fields and one is left with $24+24$ independent degrees of freedom. 
The independent fermionic fields are $\psi_{\mu}^{i}$ and $\chi^{i}$
The supersymmetry variations of the gravitini are 
\be\label{gravitinovar}
\delta\psi^i_\mu  =  2 \CD_{\mu} \epsilon^i
+\CV_\mu^i{}_j \, \epsilon^j-\frac{1}{4}\sigma^{\rho\nu} \, T_{\rho\nu}^{ij} \, \gamma_\mu \epsilon_j 
-\gamma_\mu \eta^i\, . 
\ee
In this paper, we shall use the following conventions as in \cite{Mohaupt:2000mj}. 
The symbol $D_{\mu}$ is the covariant derivative 
with respect to all the superconformal transformations, while the symbol 
$\CD_{\mu}$ is the covariant derivative 
with respect to all the superconformal transformations except the special conformal 
transformations and the fermionic $Q$ and $S$ transformations. 
The symbol $\nabla_{\mu}$ denotes the spacetime covariant derivative.
We have the definitions: 
\be
\sigma_{\mu\nu}=\frac{1}{4}[\gamma_\mu,\gamma_\nu] \, ,\qquad 
T_{\mu\nu}{}^{ij}=\frac{1}{2}T_{\mu\nu}^{-}\varepsilon^{ij} \, .
\ee
For an antisymmetric tensor field $A_{\mu\nu}$, 
\be
A^{\pm}_{\mu \nu} \equiv \half \big(A_{\mu \nu} \pm i \wt A_{\mu \nu} \big)\, , 
\ee
where $\wt A_{\mu\nu}$ is the Hodge dual of $A_{\mu\nu}$.

The supersymmetry variation of the other independent fermionic field $\chi^{i}$ contains 
a term linear in the auxiliary field $D$, and setting the variation of $\chi^{i}$ to zero is 
therefore equivalent to defining the BPS value of the field $D$. 
We shall discuss the field $D$ again in some detail below. 

The \emph{vector multiplet} is made up of 
\be\label{Vectorfields}
{\bf X}^{I} = \left( X^{I}, \O_{i}^{I}, A_{\mu}^{I}, Y^{I}_{ij}  \right) \, , 
\ee
where $X^{I}$ is a complex scalar, the gaugini $\O^{I}_{i}$ are an $\SU(2)$ 
doublet of chiral fermions, $A^{I}_{\mu}$ is a vector field, and $Y^{I}_{ij}$ are an $\SU(2)$ triplet of 
auxiliary scalars. 
The supersymmetry variation of the gaugini is 
\be\label{gauginovar}
\delta \Omega_i^I=2\gamma^\mu D_{\mu} X^I  \epsilon_i+Y^I_{ij} \, \epsilon^j
+\sigma^{\mu\nu}\mathcal{F}^{I-}_{\mu\nu}\varepsilon_{ij} \, \epsilon^j +2X^I\eta_i\, , 
\ee
with 
\be
\mathcal{F}_{\mu\nu}^{I} \equiv F_{\mu\nu}^{I} - 
\Big( \ve_{ij} \overline{\psi}^{i}_{[\mu} \g_{\nu]} \O^{j I} 
+ \ve_{ij} \overline{X}^{I}  \overline{\psi}^{i}_{[\mu} \psi^{j}_{\nu} 
+ \frac{1}{4}\overline{X}^I T_{\mu\nu}^{ij} \, \varepsilon_{ij} 
+ {\rm h.c.} \Big) \, . 
\ee

The components of the \emph{hypermultiplet} are
\begin{equation}\label{hyper}
({\cal A}^\alpha_i,\zeta^\alpha),
\end{equation}
where the scalars ${\cal A}^\alpha_i$ are pseudo-real and $\zeta^\alpha$ is a
symplectic Majorana spinor. The indices $\alpha=1\cdots 2r$ (in our case $r=1$) 
label the fundamental representation of $\USp(2r)$. 
The supersymmetry variations of the hyperini are 
\begin{equation}\label{eaaa}
\delta\zeta^\alpha=\gamma^\mu D_\mu \CA_i^\alpha\epsilon^i+\CA_i^\alpha \eta^i \, . 
\end{equation}

The first step will be  to fix the gauge symmetries 
that are external to the physical supergravity theory. In addition to the Poincar\'e supersymmetries, 
the conformal supergravity theory admits dilatations, special conformal transformations, 
a $\U(1)$ and an $\SU(2)$ rotation as local bosonic symmetries, and the conformal supersymmetries 
($S$-transformations) as local fermionic symmetries. 
We first fix the special conformal transformations by setting the dilatational gauge 
field to zero, $b_\mu=0$ (``K-gauge''). 
The  $\SU(2)$ symmetry is gauge fixed by setting 
\be\label{CAphi}
{\cal A}_\a^i= e^{-\vp} \, \delta_\a^i \, . 
\ee

The local conformal supersymmetry transformations are generated by the spinor $\eta_{i}$
in the variations above. We fix them by setting the variation of  the hypermultiplet fermions 
to zero 
\begin{equation}\label{eaaa}
0 = \delta\zeta^\alpha=\gamma^\mu D_\mu \CA_i^\alpha\epsilon^i+\CA_i^\alpha \eta^i \, . 
\end{equation}
Using~\eqref{CAphi}, we get a relation between the spinors $\eta^{i}$ and $\eps^{i}$:
\be\label{etaepsrel}
\eta^{i} = \g^{\mu} D_{\mu} \vp \, \eps^{i} \, . 
\ee
Since we have already set $b_\mu=0$, the covariant derivative only contains the 
$SU(2)$ gauge field $\CV_{\mu}^{ij}$. 

%

The field $D$ appears in the right hand side of the supersymmetry variation of the field
$\chi^{i}$, and as mentioned above, the $\delta \chi^{i}=0$ condition can be used to solve 
for the BPS value of~$D$. The field~$D$ does not appear in the right hand side of any other 
supersymmetry variation, so there is no extra condition of consistency. Moreover, the field~$D$ only 
appears in the action as a Lagrange multiplier, and therefore it is natural to impose 
its equation of motion, which gives \cite{deWit:1980tn}
\be
\vp = \CK/2 \, , 
\ee 
where the scalar field $\CK$ is defined using the prepotential $F(X)$ of the theory as:
\be\label{EminK}
e^{-\CK} :=  -i(X^I \bar{F}_I   - \bar{X}^I F_I) \, , \qquad \text{with} \quad F_{I} \equiv \frac{\p F}{\p X^{I}} \, .
\ee 
In the gauge-fixed on-shell theory, the field $\CK$ is identified with 
the Kahler potential.

For ease of presentation, we shall set the gauge fields~$A_{\mu}$ and~$\CV^{ij}_{\mu}$
to zero in the following two sections. In \S\ref{gaugeanal}, we shall reinstate them and
discuss the corresponding change in the analysis.
Our problem now reduces to solving the following gravitini variation equations:
\be\label{gravitinovaragain}
0=\delta\psi^i_\mu   =  2\big(\p_\mu-\frac{1}{2}\omega_\mu^{ab}\sigma^{ab}\big)\epsilon^i
-\frac{1}{4}\sigma^{\rho\nu} \, T_{\rho\nu}^{ij} \, \Gamma_\mu \epsilon_j 
-\gamma_\mu \eta^i\, , 
\ee
and the gaugini variation equations:
\be\label{gauginovaragain}
0=\delta \Omega_i^I  =  2\gamma^\mu \p_\mu X^I  \epsilon_i+Y^I_{ij} \, \epsilon^j
+\sigma^{\mu\nu}\mathcal{F}^{I-}_{\mu\nu}\varepsilon_{ij} \, \epsilon^j 
+2X^I  \eta_i\, , 
\ee
with $\eta^{i}$ determined in terms of $\eps^{i}$ from 
\eqref{etaepsrel}, and 
\be
\mathcal{F}_{\mu\nu}^{I} \equiv F_{\mu\nu}^{I} - 
\big(\frac{1}{4}\overline{X}^I T_{\mu\nu}^{ij} \, \varepsilon_{ij} 
+ {\rm h.c.} \big) \, . 
\ee


So far, we have not fixed the local dilatation and the $\U(1)$ gauge symmetries. 
The dilatation gauge freedom can be fixed by imposing a condition on the $\nv+1$
scalar fields $\{X^{I}\}$ to eliminate one degree of freedom. 
In the usual treatments, one imposes the condition 
\be\label{EminKone}
e^{-\CK} = 1 \, .
\ee
which is invariant under the symplectic transformation rotating the $\nv+1$ variables. 
In the full functional integral, the independent field to be integrated over are the metric variables, 
the $\nv$ scalars, and their superpartners. The reduction from $\nv+1$ scalars to 
$\nv$ scalars can be achieved by solving the condition \eqref{EminKone} to eliminate 
one of the scalar fields in terms of the others.
In this gauge, $\vp = \text{const}$, and therefore the relation \eqref{etaepsrel} reduces to~$\eta^{i}=0$.

In \cite{Dabholkar:2011ec}, an alternative way to gauge fix the local dilatation symmetry was proposed. 
Noting that the field $e^{-\CK}$ couples to the conformal mode of the metric, and 
can therefore be identified with the conformal compensator, one can equivalently keep the $\nv+1$
fields free to fluctuate in the functional integral, and constrain the conformal mode of the 
metric to be everywhere equal to a constant determined by the asymptotic boundary conditions:  
\be\label{detgcond}
\det{g} = \det{g_{\rm asymptotic}} \, .
\ee
In our case, the asymptotic metric will be that of $AdS_{2} \times S^{2}$. 
This way of gauge fixing has the advantage that the symplectic covariance 
(or the duality symmetry in string theory) is manifestly preserved in the quantum theory.
However, the field $\vp$ is now spacetime dependent and therefore a non-trivial value 
of $\eta^{i}$ determined by \eqref{etaepsrel} goes into the BPS equations 
\eqref{gravitinovaragain}, \eqref{gauginovaragain}.

It is useful to explicitly work out the map between these two gauges in order 
to pick a good starting point for the BPS equations. In the gauge  \eqref{EminKone}, 
wherein $\eta^{i}=0$, and the BPS equations \eqref{gravitinovaragain} has the same structure 
as the Killing spinor equations of supergravity, while they have an extra term 
in the gauge \eqref{detgcond}. However, this term can be absorbed into the other 
fields of the problem. More precisely, a local scaling
\be
g_{\mu\nu} \to e^{-2\vp} g_{\mu\nu} \, , \quad X^{I} \to e^{\vp} X^{I} \, , 
\quad \eps \to e^{-\vp/2}  \eps \, , \quad T_{\mu\nu} \to e^{\vp} T_{\mu\nu} 
\ee
leave the equations \eqref{gravitinovaragain}, \eqref{gauginovaragain} invariant.
It is easy to see that these transformations are nothing but the local scaling 
transformations of the conformal supergravity, and we are back to the 
first gauge condition. 

Thus we see that it is most convenient to solve the equations with $\eta^{i}=0$, 
with arbitrary fluctuations of the metric, and $\nv+1$ vector 
fields $X^{I}$ with the constraint $e^{-\CK} =1$. 
In functional integral, we should integrate over the conformal mode of the metric
and $\nv +1$ vector multiplets constrained by this one condition. 
Equivalently
we can integrate over  $\nv +1$ unconstrained vector multiplets, with the 
determinant of the metric being fixed. One should, of course, use the appropriate 
measure of the functional integral in these variables, this was done in~\cite{Dabholkar:2010uh}.

With the above gauge choices and equations of motion, the independent bosonic variables left are the metric,
the field $T_{\mu\nu}$ in the gravity sector, and the gauge field and the fields $Y^{I}_{ij}$ 
in the vector multiplet sector. 
In these variables, the gravitino variation condition is:
\be\label{gravitinovarnew}
0=\delta\psi^i_\mu  =  2\big(\p_\mu-\frac{1}{2}\omega_\mu^{ab}\sigma^{ab} \big)\, \epsilon^i
-\frac{1}{4}\sigma^{\rho\nu} \,T_{\rho\nu}^{ij} \, \Gamma_\mu \,\epsilon_j \, ,
\ee
and the gaugino variation condition is:
\be\label{gauginovarnew}
0= \delta \Omega_i^I=2\gamma^\mu \p_\mu X^I\epsilon_i+Y^I_{ij}\epsilon^j
+\sigma^{\mu\nu} \CF_{\mu\nu}^{I-}
\varepsilon_{ij} \, \epsilon^j\, .
\ee
By the above discussion, we will solve these equations without any constraints on the fields 
in the interior of the geometry. After finding the most general solutions, we can 
constrain the metric in the functional integral by \eqref{detgcond}, and keep the $\nv+1$
vector fields unconstrained.

Finally, the boundary conditions of our problem are determined 
by the classical attractor values  \cite{Mohaupt:2000mj} which we now briefly summarize.
In the Weyl multiplet, the asymptotic non-zero fields are $T^-_{rt}= 4$ and 
the metric is $AdS_{2} \times S^{2}$: 
\be \label{metricatr}
ds^2= 
\left[(r^2-1)du^2+\frac{dr^2}{r^2-1}\right] +  \left[ d\psi^2 + \sin^2\psi \, d \phi^2 \right] \, , \qquad 
 0 \leq u < 2\pi \, . 
\ee
Note that the overall scale of the metric can also be absorbed into the combination $e^{-K}$, we have 
set this to be one above. 
In the vector multiplet sector, the asymptotic values of the scalar fields, auxiliary fields, and the flux are constant:
\bea \label{sol1}
F^I_{rt} = e^{I}_{*}, \qquad F^I_{\psi\phi}=  p^I \, \sin\psi, \qquad 
X^I = X_{*}^I \, ,\qquad  Y^I_{ij}=0 \ . 
\eea 
The electric fields $e^{I}_{*}$ are determined by the real part of the scalar fields
\be  \label{sol3a} 
\bar X_{*}^I + X_{*}^I = e^I_{*} \, , 
\ee
and the constant values of the scalar fields are determined  in terms of the 
charges $(q,  p)$ by the attractor equations (here the $F_{I}$s are considered
functions of $X^{I}_{*}$):
\be \label{scalarattval}
i ( \bar X_{*}^I -  X_{*}^I) =  p^I \, ,\qquad   i ( \bar F_I -  F_I)  =  \, q_I \, . 
\ee

\section{Weyl multiplet \label{Weyl}}

As mentioned in the introduction, one advantage of the off-shell formalism is that 
we can solve the BPS equations of the Weyl multiplet sector and those of the 
vector multiplet sectors independently. In this section, we shall find all the solutions 
to the gravitino variation~\eqref{gravitinovarnew} with the constraint~\eqref{detgcond} 
and the boundary conditions~\eqref{metricatr}.  
Following the method developed in \cite{Gauntlett:2002nw}, 
our strategy to solve these equations will be to assume the existence of at least one 
spinor which solves the BPS equation. We then form spinor bilinears from the gravitini,
and then formulate algebraic and differential conditions on these bilinears using the 
spinor structure and the BPS equations. It turns out that it is much easier to solve the 
equations for the bosonic quantities. 

As mentioned in the introduction, the closely related on-shell problem (four-dimensional $\CN=2$ 
supergravity coupled to vector multiplets) has been analyzed exhaustively in \cite{Meessen:2006tu}, 
and we shall follow that treatment as far as it takes us. 
However, we still need to do some more work since, unlike \cite{Meessen:2006tu}, 
we shall keep the fluctuations of the auxiliary fields. Another difference is that our application 
to the functional integral dictates an analytic continuation to Euclidean space,
which will further constrain the space of BPS solutions.
Due to the off-shell nature, and the analytic continuation, the space of solutions 
that we find is different from that of \cite{Meessen:2006tu}, neither one is a subset of the other.

We shall do the first part of the analysis in Minkowski space, and we shall indicate below 
when we do the analytic continuation to Euclidean space. 
For technical ease, (especially with the Fierz rearrangements that we shall use repeatedly),  
we shall work with the Dirac spinor $\psi$ formed out of the two chiral-Majorana spinors 
$\eps_{i}$ as $\psi=\epsilon^1+i\epsilon_2$, in terms of which the BPS equation \eqref{gravitinovarnew} 
is written as\footnote{Here we have used the fact that for a positive chirality spinor $\eps^{i}$ and 
any antisymmetric tensor $A^{\mu\nu}$, one has 
$A^{\mu\nu} \s_{\mu\nu} \eps^{i} =  A^{\mu\nu-} \s_{\mu\nu}  \eps^{i}$.}:  
\be \label{bpspsi}
\big(\p_\mu-\frac{1}{2}\omega_\mu^{ab}\sigma^{ab} \big)\psi
+\frac{i}{16}\sigma^{\rho\nu} \, T_{\rho\nu} \, \Gamma_\mu\psi = 0 \, .  
\ee
This equation is inherently complex. If we impose for example a Weyl or Majorana 
condition on the spinor $\psi$, then we do not get any non-trivial solutions. 
We see that, although we assumed the existence of one real supercharge, 
the structure of the theory gives two real supercharges. In the following, we 
shall see that we have four real supercharges that are preserved in our 
solution\footnote{This is also the case in the Lorentzian on-shell analysis  \cite{Meessen:2006tu}, and seems to be 
a feature of the BPS equations of the theory \cite{Batrachenko:2004su}.}.

We define the following bilinears constructed out of $\psi$
\be
f_1=\bar\psi\psi \, , \qquad f_2=i\bar\psi\Gamma_5\psi \, , \qquad K_\mu=i\bar\psi\Gamma_\mu\psi \, , 
\ee
which are a spacetime scalar, a pseudo-scalar, and a vector, respectively. 
For future use, we form a complex scalar field $X$ and define its magnitude and phase by:
\be \label{defRth}
f_{1} + i f_{2} =: X \equiv R \, e^{i \Theta} \, . 
\ee
The BPS equation \eqref{bpspsi} leads to the following first order differential equations: 
\bea \label{pXXbar}
&& \p_\mu X =  \frac{1}{4}T^{-}_{\mu\nu} \, K^\nu \, ,  \qquad 
\p_\mu \bar X =  \frac{1}{4} T^{+}_{\mu\nu} \, K^\nu \, ,  \label{partialXmu}  \\
&& \qquad \nabla_\mu K_\nu =  -\frac{1}{8}T^{+}_{\mu\nu} \, X - \frac{1}{8} T^{-}_{\mu\nu} \, \bar X  \, . \label{Kmudiff}
\eea
Since the RHS of \eqref{Kmudiff} is manifestly antisymmetric in $\mu \leftrightarrow \nu$, we have:
\be\label{Killvec}
\nabla_\mu K_\nu + \nabla_\nu K_\mu  = 0 \, , 
\ee
that is, $\wh K \equiv K^{\mu} \p_{\mu}$ is a Killing vector of the geometry. 
We shall assume that this vector is everywhere timelike, and use it to define the coordinate $t$ by:
\be 
\wh K = \frac{\p}{\p t} \, . 
\ee
By contracting equation \eqref{partialXmu} with $K^{\mu}$, and using the antisymmetry of  $T_{\mu\nu}$,
we deduce that 
\be
 \p_{t} X= \p_{t} \bar X = 0 \, . 
 \ee
Using Fierz identities to rearrange four fermion terms \cite{Caldarelli:2003pb, Nieves:2003in}, 
we find the following algebraic relation between these quantities:
\be
K^\mu K_{\mu}=-R^{2} \, .
\ee
Using equation \eqref{partialXmu}  we can determine $T_{\mu\nu}$ to be:
\be \label{Tsol}
T_{\mu\nu}=\frac{8}{R^{2}}[(K_\mu\p_\nu f_1-K_\nu\p_\mu f_1)-K^\rho
\p^\s f_2 \, \varepsilon_{\s\rho\mu\nu}] \, .
\ee

We now define the three linearly independent bilinears (with $\psi=\epsilon^1+i\epsilon_2$)
\be
\Phi^{(1)}_\mu =i \, \bar \eps_{2} \g_{\mu} \eps^{1} \, + \,  i \, \bar \eps_{1} \g_{\mu} \eps^{2} \, , \qquad 
\Phi^{(2)}_\mu =\bar \eps_{2} \g_{\mu} \eps^{1} \, -  \bar \eps_{1} \g_{\mu} \eps^{2} \, , \qquad 
\Phi^{(3)}_\mu = i \, \bar \eps_{1} \g_{\mu} \eps^{1} - i \, \bar \eps^{2} \g_{\mu} \eps_{2} \, . 
\ee
These obey the algebraic relations 
\be
K^{\mu} \Phi_{\mu}^{(\a)} =0 \, , \qquad \Phi^{(\a)}{}^{\mu} \, \Phi^{(\b)}_{\mu}  = R^{2} \, \delta^{\a\b} \, , \quad \a, \b=1,2,3 \, .
\ee
The BPS equation \eqref{bpspsi} implies that one-forms 
$\Phi^{(\a)} \equiv \Phi^{(\a)}_{\mu} d x^{\mu}$
are closed, {\it i.e.} they obey the differential equations:
\be\label{oneform}
d \Phi^{(\a)} =0 \, . 
\ee
We can therefore choose local coordinates $y^{\a}$,  
such that $\Phi^{(\a)} = d y^{\a}$, $(\a=1,2,3)$. Without loss of generality, 
we can also choose these coordinates to be mutually orthogonal.

From the above discussion, we deduce that the metric takes the form 
\be\label{metricform}
ds^2=-R^2 \big(dt+V \big)^2+ \frac{1}{R^2} \big(\sum_{\a=1}^{3} dy^{\a} dy^{\a} \big) \, . 
\ee
The one-form $V$ can be chosen to have the form $V =V_\a dy^\a$ (by a reparameterization of 
the coordinate $t$). Further, since $\p_{t}$ is a Killing vector, 
\be
\p_{t} R =0 \, , \qquad \p_{t}V=0 \, . 
\ee
The BPS equation implies as usual a differential condition on the one-form 
$K \equiv R^{2}(dt+V)$, which translates into a condition on $V$ (with $\ve_{0123} =1$):
\be
\p_\rho V_\sigma-\p_\sigma V_\rho=\frac{2}{R^4} \, K^\mu \, 
\p_\nu\Theta \, \varepsilon^\nu{}_{\mu\rho\sigma} \, ,
\ee
where $\Theta$ was defined in \eqref{defRth}.
Since $V_{t}=0$, and $\p_{t} V = \p_{t} X=0$, we get an equation in the three 
dimensional space $\{y^{\a}\}$:
\be \label{dVeqn}
\p_\a V_\b-\p_\b V_\a=-\frac{2}{R^2}  \, \p^\g\Theta \, \varepsilon_{\g\a\b} \, , 
\ee
where now, we use the three dimensional flat metric to raise and lower the indices. 
The equation \eqref{dVeqn} has an associated integrability condition:
\be\label{integrabilityV}
0=\p_\a \big(\frac{1}{R^2}  \, \p_\d\Theta \, \varepsilon^\d{}_{\b\g} \big)
dx^\a\wedge dx^\b \wedge dx^\g \, ,
\ee
which can be rewritten as:
\be\label{integrabilityV}
0=\p^\a \big(\frac{1}{R^2}  \, \p_\a\Theta  \big) \, .
\ee

This finishes the general analysis of the algebraic and differential conditions on the 
spinor bilinears. One can try to construct higher tensors with two or more legs, but 
these turn out to be determined algebraically in terms of the vectors 
and one-forms. One therefore does not get any new differential constraints for the 
bilinear fields. Some explicit relations are written e.g.~in \cite{Caldarelli:2003pb}. 
We present one such example in Appendix~\ref{Mabanal}, which will be useful to us later.

To proceed towards the quantum entropy function, we change variables 
in the flat three dimensional space spanned by $y^{\a}$ to spherical-polar coordinates:
\be\label{genmet}
ds^2=-R^2 \big(dt+V \big)^2+ \frac{1}{R^2} \big(d\rho^{2} + \rho^{2} (d\psi^{2} + \sin^{2} \psi \, d\phi^{2}) \big) \, .
\ee
According to the discussion in \S\ref{bpseqns}, we should impose the gauge condition that the 
determinant of the metric times a function of the scalars is fixed in terms of the asymptotic 
$AdS_{2} \times S^{2}$ space which has $R=\rho$. This gives us:
\be \label{Reqrho}
R(y^{\a}) = \rho \, e^{2 \vp}. 
\ee

At this point, it seems like we can use the $U(1)$ gauge symmetry (under which 
the field $\Theta$ translates) to set it to zero. However, when we include 
the $U(1)$ gauge field, we would like to use the gauge freedom to fix a different 
field, so we would like to do the analysis here without using the gauge condition. 
The field $\Theta$ obeys the second order differential constraint \eqref{integrabilityV}. 
It turns out that after analytic continuation to Euclidean space, and 
imposing  the $AdS_{2}$ boundary conditions, the only solution to this equation
is constant~$\Theta$. We present the details of this statement in Appendix \S\ref{smoothV}.

We are led to the final conclusion that the BPS configurations in the Weyl multiplet 
sector are all conformally equivalent to $AdS_{2} \times S^{2}$. They are 
labelled by one real function $\vp(X^{I}(x^{\mu}))$ of the scalars which parameterizes the 
conformal mode of the metric. 
The field $T_{\mu\nu}$ is determined to be a constant, with $T^{-}_{rt}$ determined 
by the  constant $\Theta$:
\be
T^{-}_{rt} = 4 \, e^{i\Theta} \, . 
\ee

Now we do an analytic continuation to Euclidean space following \cite{Sen:2008vm}. 
Asymptotically, we have the $AdS_{2} \times S^{2}$ metric 
\be\label{ads2s2}
ds^{2} = \left[-\rho^{2} dt^{2} + \frac{d \rho^{2}}{\rho^{2}}\right] + \left[d\psi^{2} + \sin^{2} \psi \, d\phi^{2}\right]  \, . 
\ee
We shall keep the $S^{2}$ part of the metric as it is. On the $AdS_{2}$ part, we begin 
by analytically continuing $t\to -i\tau$ to get
\be \label{eec1}
ds^2  = \rho^2 d\tau^2+\frac{d\rho^2}{\rho^2} \, . 
\ee
We then introduce new coordinates $(\eta,\theta)$ through:
\be \label{eseries}
z= \tau + i \, \rho^{-1}, \quad w= (1+iz)/(1-iz), \quad
\tanh(\eta/2) \, e^{i\theta} = w\, .
\ee
These coordinate changes map the Euclidean $AdS_2$ represented as an upper half plane
in the complex coordinate $z$ to the interior of a unit hyperbolic disk described 
by the complex coordinate $w$. 
The variables $(\tanh{1\over 2}\eta, \theta)$ are the usual polar coordinates on the unit disk in the $w$-plane. 
In the $(\eta, \theta)$ coordinates the
solution \eqref{ads2s2} becomes 
\be \label{eads2s2}
ds^2 = \left[ d\eta^2 + \sinh^2\eta \, d\theta^2 \right]  + \left[d\psi^{2} + \sin^{2} \psi \, d\phi^{2}\right]  \, . 
\ee

We can now solve for the explicit form of the Killing spinor. We use the following gamma matrix representation 
in terms of the Pauli matrices:
\begin{equation}
\gamma_{0}=  \sigma_1\otimes 1\ , \quad \gamma_{1}=  \sigma_2\otimes
1\ , \quad \gamma_{2}=  \sigma_3\otimes \sigma_1\ , \quad 
\gamma_{3}=  \sigma_3\otimes \sigma_2 \ .
\end{equation}

From the discussion on gauge-fixing 
in \S\ref{bpseqns}, it follows that we can solve the Killing spinor equation on the $\vp={\rm const}$ configuration,
which is pure $AdS_{2} \times S^{2}$ space, and the constant value of $T$ as determined above.
We briefly present this analysis below, following~\cite{Dabholkar:2010uh}. 
In the Euclidean theory, we should use symplectic Majorana spinors. This is  
achieved by defining the spinors $\xi^{i}_{\pm}$, ($i=1,2$, and the $\pm$ subscripts denote the chirality):
\be 
 \epsilon_i=i\ve_{ij}\xi^j_{-} \, , \qquad  \epsilon^i=\xi^i_{+}  \, ,
\ee
that obey the symplectic Majorana condition:
\begin{equation} \label{Majsymp}
(\xi^{i}_{\pm})^{*}=-i \ve_{ij} \, (\sigma_1\otimes \sigma_2) \,\xi^{j}_{\pm} \, .
\end{equation}

The Killing spinor equation is solved for an unconstrained 
Dirac spinor $\xi^i=\xi^i_{+}+\xi^i_{-}$, obtained by double the space (and 
then imposing the constraint \eqref{Majsymp} at the end). 
We represent the Dirac spinor $ \xi$ as a direct product $ \xi= \xi_{AdS_2}\otimes \xi_{S^2}$ 
where $ \xi_{AdS_{2}}$ and $ \xi_{S^{2}}$ are two component spinors.
The Killing spinor equations \eqref{bpspsi} take the diagonal form ($\mu=0,1$, $j=2,3$): 
\begin{eqnarray}
 D_{\mu} \, \xi_{AdS_2} & = & \frac{i}{2}(\sigma_3\otimes 1) \, \gamma_{\mu} \,  \xi_{AdS_2} \ , \\
 D_{j} \, \xi_{S^2} & = & \frac{i}{2}(\sigma_3\otimes 1) \, \gamma_{j} \,  \xi_{S^2} \ . \\
\end{eqnarray}
The $AdS_{2} \times S^{2}$ space is maximally supersymmetric and we find four complex Killing 
spinors on this space. The solutions are
\begin{eqnarray} \label{Killspin1}
 \xi^i_{--}= e^{-\frac{i}{2}(\theta+\phi)}\left(\begin{array}{c}
\cosh\frac{\eta}{2}\cos\frac{\psi}{2}\\
                                                         \sinh\frac{\eta}{2}
\cos\frac{\psi}{2}\\
                                                        
-\cosh\frac{\eta}{2}\sin\frac{\psi}{2}\\
                                                        
-\sinh\frac{\eta}{2}\sin\frac{\psi}{2}\end{array}\right)  & \ , \qquad & 
 \xi^i_{-+} = e^{-\frac{i}{2}(\theta-\phi)}\left(\begin{array}{c}
\cosh\frac{\eta}{2}\sin\frac{\psi}{2}\\
                                                         \sinh\frac{\eta}{2}
\sin\frac{\psi}{2}\\
                                                        
\cosh\frac{\eta}{2}\cos\frac{\psi}{2}\\
                                                        
\sinh\frac{\eta}{2}\cos\frac{\psi}{2}\end{array}\right) \ , \nonumber \\
 \xi^i_{+-} = e^{\frac{i}{2}(\theta-\phi)}\left(\begin{array}{c}
\sinh\frac{\eta}{2}\cos\frac{\psi}{2}\\
                                                         \cosh\frac{\eta}{2}
\cos\frac{\psi}{2}\\
                                                        
-\sinh\frac{\eta}{2}\sin\frac{\psi}{2}\\
                                                        
-\cosh\frac{\eta}{2}\sin\frac{\psi}{2}\end{array}\right) & \ , \qquad &  
 \xi^i_{++} = e^{\frac{i}{2}(\theta+\phi)}\left(\begin{array}{c}
\sinh\frac{\eta}{2}\sin\frac{\psi}{2}\\
                                                         \cosh\frac{\eta}{2}
\sin\frac{\psi}{2}\\
                                                        
\sinh\frac{\eta}{2}\cos\frac{\psi}{2}\\
                                                        
\cosh\frac{\eta}{2}\cos\frac{\psi}{2}\end{array}\right)  \, . 
\end{eqnarray}

\section{Vector multiplets \label{vectors}}

In this section, we shall find the solutions in the vector multiplet sector. The results in 
this section were already found in  \cite{Dabholkar:2010uh}, here we rederive them using the spinor 
bilinear method. The two methods are almost equivalent but have small differences,
it may be useful to keep both in our toolkit for future problems. 
As in the Weyl multiplet sector, we look for solutions that preserve 
at least one supercharge. We shall find that the most general BPS solutions 
actually preserve four supercharges. This was conjectured to be true for 
off-shell localization~\cite{Banerjee:2009af}, and our work provides a proof 
within the context that we work in.

The BPS equations in the vector multiplet sector \eqref{gauginovarnew}: 
\be\label{gauginovarnewag}
0= \delta \Omega_i^I=2\gamma^\mu \p_\mu X^I\epsilon_i+Y^I_{ij}\epsilon^j
+\sigma^{\mu\nu} \CF_{\mu\nu}^{I-}
\varepsilon_{ij} \, \epsilon^j\, .
\ee
The metric, the field $T_{\mu\nu}$, and the spinor $\eps$ are the solutions 
to the Weyl multiplet equations as derived in \S\ref{Weyl}. 
In particular, $\eps$ is a Killing spinor of the $AdS_{2} \times S^{2}$
background, and the supersymmetry variation in the above equation is 
with respect to the linear combination of supercharges 
$Q_{1}$ obeying $Q_{1}^{2} = 4(L_{0}-J_{0})$.

In solving these equations, it is important to be careful about the analytic continuation 
that we perform to go to Euclidean space. Apart from the analytic continuation of the 
metric and the spinors as in \S\ref{Weyl}, we must also perform a continuation of 
the fields $X^{I}, \bar X^{I}$. The solutions of the equations 
\eqref{gauginovarnewag} depend quite crucially on this analytic continuation. 
The correct choice our application turns out to be the one used in \cite{Cortes:2003zd}, 
the scalars $X, \overline X$ in a vector multiplet are taken to be two independent 
real scalars in this treatment. 
Note that the analytic continuation in question depends quite 
crucially\footnote{Our analytic continuation is similar to the one used in \cite{Pestun:2007rz}.
A different choice of continuation gives a different (and less restrictive) set of 
solutions to the BPS equations. 
If we do the analytic continuation on vector multiplets so that 
$\Sigma ^{I} = H^{I} + i J^{I} \, , \, \bar \Sigma^{I} = H^{I} - i J^{I} $,
we find that in addition to the solutions \eqref{vecsols}, we get a new family 
of solutions which are parameterized by the function $J$ that is unconstrained 
by supersymmetry. Imposing the Bianchi identities on $F_{ab}$ give differential 
constraints on $J$, which do admit non-trivial solutions.}
on the physical application at hand\footnote{We thank Jo\~ao Gomes for many discussions
about the correct analytic continuation. Note that a previous version of  \cite{Dabholkar:2010uh}
had an error regarding this issue.}. Our choice is fixed by demanding that the action 
of the fluctuations inside the functional integral of localization be bounded 
below.
An example of a different consistent choice is the computation 
of the logarithmic corrections to the classical entropy \cite{Banerjee:2010qc, Banerjee:2011jp}. 
The difference can be traced to the fact that the localization computation
involves deforming the physical action by a $Q$-exact term. As the deformation 
parameter becomes larger, it is the latter action which needs to have a positive 
definite fluctuation.

The BPS equations of the  Euclidean theory are:
\begin{eqnarray}\label{Qvars}
 0 && =\frac{1}{2}(F_{\mu\nu}^{I-}-\frac{1}{4}\bar{X}^{I} \, T^{-}_{\mu\nu}) \, 
 \gamma^{\mu} \, \gamma^{\nu} \, \xi^{i}_+ +2i \displaystyle{\not}\partial X^{I} \, \xi^i_-+Y^{Ii}_j \,  \xi^j_+ \ , \\
 0 && =\frac{1}{2}(F_{\mu\nu}^{I+}-\frac{1}{4}X^{I} \, T^{+}_{\mu\nu}) \, 
 \gamma^{\mu} \, \gamma^{\nu} \,  \xi^{i}_- +2 i \displaystyle{\not}\partial \bar{X}^{I} \,  \xi^i_+ +Y^{Ii}_j \,  \xi^j_- \ .
\end{eqnarray}
We can, as before, add these two equations to write two equivalent equations for the 
Dirac spinor and its conjugate. 

We now consider the fluctuations of the various fields away from the attractor values 
\eqref{sol1}, \eqref{sol3a}, \eqref{scalarattval}, and label the various fluctuations as follows:
\be\label{offshellpar}
F^{ab} = F^{ab}_{*} + f^{ab} \, , \qquad 
X^{I} := X_{*}^{I } + \Sigma^{I}\, , \qquad \bar X^{I} := \bar X_{*}^{I } + \bar \Sigma^{I} \, ,
\ee
\be\label{defHJ}
\text{with} \qquad \Sigma ^{I} = H^{I} + J^{I} \ , \qquad \bar \Sigma^{I} = H^{I} - J^{I} \ .
\ee
Here we have used the Euclidean continuation discussed above. 
Similarly, we have the analytically continued auxiliary fields 
$Y_{11}= -i K_{2}e^{i\alpha}, Y_{22} =i K_{1}e^{i\beta}, Y_{12}= Y_{21}=K_{3}$, 
with $K_{i}$ real. Here we have allowed for an arbitrary phase in the fields~$K_{1,2}$ in addition 
to the analytic continuation mentioned above, this phase will be fixed and discussed below.

Since the vector multiplet equations are decoupled, we can analyze each one of 
them separately. Suppressing the vector index $I$, 
the BPS equations for the fluctuations are:
\bea\label{1stBPSequ}
\frac{1}{2}f_{ab}\s^{ab}\xi^+_{++}+2i\gamma^\mu\p_\mu H\xi^+_{++}
-2i\gamma^\mu\p_\mu J\gamma_5\xi^+_{++}-&&2iH\s^{01}\xi^+_{++}
+2iJ\s^{23}\xi^+_{++} \cr 
&& \qquad +K_{3}\xi^+_{++} + i K_{1} e^{i\beta}\xi^-_{--} = 0
\eea
\bea\label{2ndBPSequ}
\frac{1}{2}f_{ab}\s^{ab}\xi^-_{--}+2i\gamma^\mu\p_\mu H\xi^-_{--}
-2i\gamma^\mu\p_\mu J\gamma_5\xi^-_{--}-&& 2iH\s^{01}\xi^-_{--} +2iJ\s^{23}\xi^-_{--} \cr
&& \qquad -K_{3}\xi^-_{--} + i K_{2} e^{i\alpha}\xi^+_{++} = 0
\eea

As in \S\ref{Weyl}, we now use these BPS equations to write down first order 
differential equations for the bosonic fields in the vector multiplets. The primary 
equations are written in 
\eqref{1stBPSequ1}--\eqref{2ndBPSequ4} in terms of quantities called $a_{\mu}$,
$b_{\mu}$, $\widehat b_{\mu}$, $\widehat b'_{\mu}$, $M_{ab}$, $N_{ab}$, $N'_{ab}$,
which are defined in \eqref{defamu}--\eqref{Npr01}. 
Writing out the real and imaginary parts of equations 
\eqref{1stBPSequ1}--\eqref{2ndBPSequ4} separately, we get sixteen equations whose linear 
combinations can be written as follows.

For the fields $f_{ab}$, we get: 
\bea
\label{fSol.1} && -f_{01}+\sinh\eta\sin\psi f_{12}-\cosh\eta\cos\psi f_{23} = 0 \, , \\
\label{fSol.2} && -\cosh\eta\cos\psi f_{01}+\sinh\eta\sin\psi f_{03}-f_{23} = 0 \, ,  \\
\label{fSol.3} && f_{03}+f_{12}\cosh\eta\cos\psi+f_{23}\sinh\eta\sin\psi + \cr 
&& \qquad \qquad  \qquad  \qquad 
\half \big( K_{1}\cos(\theta+\phi-\beta) + K_{2}\cos(\theta+\phi+\alpha) \big) \cosh \eta= 0 \, ,  \\ 
\label{fSol.5} && f_{12}+f_{03}\cosh\eta\cos\psi+f_{01}\sinh\eta\sin\psi + \cr
&& \qquad \qquad   \qquad  \qquad 
\half \big( K_{1} \cos(\theta+\phi-\beta)+ K_{2}\cos(\theta+\phi+\alpha) \big)  \cos \psi = 0 \, ,  \\
\label{fSol.4} && -f_{02}\cos\psi+f_{13}\cosh\eta +
\half \big(K_1\sin(\theta+\phi-\beta)+K_2\sin(\theta+\phi+\alpha)\big)\cosh\eta= 0 \,,  \\ 
\label{fSol.6} && -f_{02}\cosh\eta +  f_{13}\cos\psi+
\half\big(K_1\sin(\theta+\phi-\beta)+K_2\sin(\theta+\phi+\alpha)\big)
\cos\psi = 0 \, . 
\eea
For the scalar fields $H$ and $J$, we get 
\be\label{Sol.1}
(\p_\theta - \p_{\phi}) H = 0 \, , \qquad  (\p_\theta - \p_{\phi}) J = 0 \, , 
\ee
which means that we can write $\p_\theta H = \half \p_{\theta+\phi} H$, and similarly for $J$.
We then have 
\bea\label{pthphiH}
\p_{\theta+\phi} H + \frac{\sin\psi\sinh\eta}{\sin^2\psi+\sinh^2\eta} \frac{K_1\cos(\theta+\phi-\beta)-K_2\cos(\theta+\phi+\alpha)}{2}\cos\psi= 0 \, , \\
\label{pthphiJ} \p_{\theta+\phi} J + \frac{\sin\psi\sinh\eta}{\sin^2\psi+\sinh^2\eta} \frac{K_1\cos(\theta+\phi-\beta)-K_2\cos(\theta+\phi+\alpha)}{2}\cosh\eta= 0 \, . 
\eea
We also have equations determining $K_{3}$ in terms of the other fields:
\bea\label{JSol.2}
-2\sinh\eta\cos\psi\p_\eta J+2\sin\psi\cosh\eta\p_\psi J-2H+2J\cosh\eta\cos\psi+K_{3}\cosh\eta = 0  \, , \\
\label{KSol}
-2\sinh\eta\cos\psi\p_\eta H+2\sin\psi\cosh\eta\p_\psi H-2H\cosh\eta\cos\psi+2J+K_{3}\cos\psi = 0 \,   .
\eea
Finally, we have two equations involving the fields $H$ and $J$ respectively, each of which 
does not involve any of the other fields:
\bea\label{JSol.1}
\cosh\eta\sin\psi \, \p_\eta J+\sinh\eta\cos\psi \, \p_\psi J-J\sinh\eta\sin\psi+ \qquad \qquad \qquad \qquad \cr
\frac14 \big(K_1\sin(\theta+\phi-\beta)-K_2\sin(\theta+\phi+\alpha)\big)\cosh\eta &=& 0 \, , \\
\label{HSol.2}
\cosh\eta\sin\psi \, \p_\eta H+\sinh\eta\cos\psi\, \p_\psi H+H\sinh\eta\sin\psi + \qquad \qquad \qquad \qquad  \cr
\frac14\big(K_1\sin(\theta+\phi-\beta)-K_2\sin(\theta+\phi+\alpha)\big)\cos\psi&=& 0 \, .
\eea

At this point, we discuss the analytic continuation of the fields~$Y^{ij}$ in more detail. We wrote out the 
BPS equations in all generality above. Our point of view is that, since we do not know the rules of quantum 
gravity very well, an analytic continuation suggested from the microscopic string theory should be 
taken seriously. In our system here, it seems that for generic~$\a,\b$, the solution manifold includes an arbitrary 
doublet of functions~$K_{\pm}$. However, if we set\footnote{Note that this analytic continuation seems to be spacetime dependent, which is certainly unusual. 
We do not have more to say about it here except that the factor arises from an inner product of two 
supercharges on a spacetime that has already been fixed completely by the gravity analysis (see~\S\ref{Weyl}). 
The situation is therefore analogous to a field theory on a fixed curved spacetime, where such a continuation 
would be slightly less unusual.}~$\b = -\a = \theta + \phi$, the solution set shrinks, 
and we recover the solution set of~\cite{Dabholkar:2010uh, Dabholkar:2011ec} that was 
consistent with the microscopic string theory. In the following analysis, we make this choice for~$\a,\b$. 
We now show that the equations (\ref{fSol.1}--\ref{HSol.2}) together with the boundary 
conditions and smoothness completely determine the fluctuations. The conditions 
are that all the fluctuating fields are smooth in the interior, and 
decay towards the boundary $\eta \to \infty$.

Firstly, one can write the equations \eqref{JSol.1} and \eqref{HSol.2} as:
\bea \label{Hagain}
\big(\coth\eta \, \p_\eta + \cot\psi \, \p_\psi -1\big)J  &=& 0 \, , \\
\label{Jagain}
\big(\coth\eta\, \p_\eta + \cot\psi\, \p_\psi +1\big)\,H &=& 0 \, .
\eea
The most general solution of equations \eqref{Hagain}, \eqref{Jagain}
and \eqref{Sol.1} is:
\be\label{HJSol}
H=f_1(v,\theta+\phi)\sqrt{\frac{\cos\psi}{\cosh\eta}},
\qquad J=f_2(v,\theta+\phi)\sqrt{\frac{\cosh\eta}{\cos\psi}},\quad (v\equiv\cosh\eta\cos\psi) \, , 
\ee 
with $f_1$ and $f_2$ being arbitrary functions of their arguments. Considering the Laurent expansion of 
$f_1(v)$ and $f_2(v)$ around $v=0$, and the smoothness and boundary condition for $H$ and $J$, 
we find that
\be \label{J0HC}
J=0  \, , \qquad H=\frac{C(\theta+\phi)}{\cosh\eta} \, , 
\ee
where $C$ is an arbitrary real function. Plugging this in \eqref{JSol.2} determines $K_{3}$ to be:
\be
K_{3} = \frac{2H}{\cosh \eta} \, .
\ee
One can check that this satisfies \eqref{KSol} automatically. Plugging in $J=0$ in 
\eqref{pthphiJ}, we find $K_{1}-K_{2}=0$. From \eqref{pthphiH}, we now get  
$C=$ constant.
%

We now look at the equations for the field strengths $f_{ab}$. 
From \eqref{fSol.4}, \eqref{fSol.6}, we get $f_{02}=f_{13}=0$.
Solving \eqref{fSol.1}--\eqref{fSol.5}, we get (using $K_{1}=K_{2}$):
\ben
f_{\theta\eta} & = & \sinh\eta f_{01}=\frac{K_{1}\sin2\psi\sinh^2\eta}{\cosh2\eta-\cos2\psi} \, , \\
f_{\theta\psi} & = & \sinh\eta f_{03}=-\frac{K_{1}\sin^2\psi\sinh2\eta}{\cosh2\eta-\cos2\psi} \, , \\
f_{\eta\phi} & = & \sin\psi f_{12}=-\frac{K_{1}\sinh^2\eta\sin2\psi}{\cosh2\eta-\cos2\psi} \, , \\
f_{\psi\phi} & = & \sin\psi f_{32}=\frac{K_{1}\sin^2\psi\sinh2\eta}{\cosh2\eta-\cos2\psi} \, . 
\een
The Bianchi identity for the field strengths $f_{ab}$ gives us: 
\be
\p_\psi f_{\theta\eta}-\p_\eta f_{\theta\psi}=0 \, . 
\ee
Plugging in the values of the fields strengths in terms of $K_{1}$, and defining the variable 
$K' \equiv \frac{K_{1}\sin\psi\sinh\eta}{\cosh2\eta-\cos2\psi}$, the Bianchi identity is reexpressed as:
\be
\big(\cot\psi \p_{\psi} + \coth\eta \p_{\eta} \big) K' =0 \, , 
\ee
whose solution is 
\be
K' = K'(v,\theta,\phi) \, , \qquad \text{$v\equiv \cosh\eta \cos \psi$ as above.}
\ee
Doing a Laurent expansion as above, and demanding smoothness in the interior and 
falloff at the boundary leaves us with only the trivial solution 
$K'=0 \Rightarrow K_{1} = K_{2} =0 \Rightarrow f_{ab}=0$.

We summarize that the most general solution of these equations are:
\ben \label{vecsols}
H^{I}= \frac{C^{I}}{\cosh\eta}  \, , \qquad  
K^{I}_{3}=\frac{2C^{I}}{(\cosh\eta)^{2}}  \, , \qquad J^{I}= K^{I}_{1,2} = f^{I}_{ab} =0 \, ,
\een
with the $C^{I}$s being arbitrary real numbers. 
We can now check that the solution \eqref{eads2s2}, \eqref{vecsols} preserves four supercharges.

\section{Inclusion of the gauge fields $A_{\mu}$ and $\CV_{\mu}^{ij}$ \label{gaugeanal}}

We performed the analysis so far with the assumption that the gauge fields $A_{\mu} = \CV^{ij}_{\mu}=0$.
In this section, we shall remove this assumption, and we shall describe how the analysis in~\S\ref{Weyl} and \S\ref{vectors} changes. The technique remains exactly the same, 
we use the BPS equations obeyed by the Killing spinors to write down first order equations for the 
bosonic quantities formed from the bilinears. 



We begin with the Weyl multiplet. 
The BPS equations~\eqref{bpspsi} change in that the partial derivative $\p_{\mu}$ is replaced 
by the covariant derivative~$D_{\mu}$, which includes a coupling to the~$U(1)$ and~$SU(2)$ 
gauge fields. The bilinears~$K_{\mu}$ are not charged with respect 
to the~$U(1)$ gauge field, so the first order differential equation~\eqref{Killvec} does not 
change, which means that $K$ remains a Killing vector.
We can therefore choose the same coordinates as before so that $\p_{t}$ is a Killing vector.

The scalar field satisfies the following equation 
\be\label{XEqu1}
D_\mu X=\frac{1}{4}T^-_{\mu\nu}K^\nu\,,\quad D_\mu \bar X=\frac{1}{4}T^+_{\mu\nu}K^\nu\,,
\ee
with 
\be
D_\mu X=\p_\mu X+iA_\mu X \, . 
\ee
Denoting the radial and phase parts of the complex field $X$ by the functions $R$ and $\Theta$ as before, and contracting~\eqref{XEqu1} with $K^\mu$, we get
\be\label{XEqu2}
K^\mu D_\mu X=0 \Rightarrow \p_t R=0,\quad \p_t \Theta+A_t=0 \, . 
\ee
Inverting \eqref{XEqu1}, we can solve for the fields~$T^{\pm}_{\mu\nu}$:
\ben
&&T^+_{\rho\sigma}=\frac{4}{R^2}D_\beta\bar X[K_\rho\delta^\beta_\sigma-K_\sigma\delta^\beta_\rho+iK_\alpha\varepsilon^{\alpha\beta}{}_{\rho\sigma}] \, , \\
&&T^-_{\rho\sigma}=\frac{4}{R^2}D_\beta X[K_\rho\delta^\beta_\sigma-K_\sigma\delta^\beta_\rho-iK_\alpha\varepsilon^{\alpha\beta}{}_{\rho\sigma}] \, . 
\een

The one forms $\Phi^{(a)}$, $a=1,2,3$, are no longer closed, instead, they satisfy the following equation
\be\label{PhiEqu2}
d\Phi^{(a)}=\varepsilon_{abc}\, B^{(b)}\wedge \Phi^{(c)} \, ,
\ee 
where 
\be
B^{(1)}=\frac{i}{2}(V^1{}_2+V^2{}_1),\quad B^{(2)}=\frac{1}{2}(V^1{}_2-V^2{}_1),\quad B^{(3)}=iV^1{}_1 \, . 
\ee
From \eqref{PhiEqu2} we get an integrability condition on the~$SU(2)$ field strength $R(V)^i{}_j$. 
Writing~$R(V)^{(a)} \equiv \sigma^{a}_{ij} \, R(V)^{ij}$ in terms of the Pauli matrices,  we have
\be
\varepsilon_{abc} \, R(V)^{(b)}\wedge\Phi^{(c)}=0 \, . 
\ee
Using the forms $\Phi^{(a)}$, we can locally define three coordinates $(y_1,y_2,y_3)$ as
\be
\Phi^{(1)}=\psi_1(t,{y_a}) \, dy_1\,, \qquad \Phi^{(2)}=\psi_2(t,{y_a}) \, dy_2\,, \qquad \Phi^{(3)}=\psi_3(t,{y_a}) \, dy_3 \, . 
\ee
The metric now can be written as
\be
ds^2=-R^2(dt+V)^2+\frac{1}{R^2}\Phi^{(a)}\Phi^{(a)} \, . 
\ee
The equation for $V$ also gets modified in that the ordinary derivative of $\Theta$ is replaced by the 
covariant derivative
\be\label{VEqu1}
dV=\frac{2}{R^2}i_k*(d\Theta+A) \, .
\ee

We can now solve~\eqref{PhiEqu2} for $B^{(a)}_\mu$ in terms of the one-forms~$\Phi^{(a)}$. 
The components along the Killing vector direction simplify:
\be \label{Bphitcomp}
\p_t \psi_{(i)}=0 \, , \qquad  B^{(i)}_t=0 \, ,
\ee
there is one algebraic relation between the gauge field components:
\be
B^{(1)}_{y_1}=-\frac{\psi_1}{\psi_3}B^{(3)}_{y_3} \, , \qquad B^{(2)}_{y_2}=-\frac{\psi_2}{\psi_3}B^{(3)}_{y_3} \, ,
\ee
and finally, the other components of~$B^{(a)}_{\mu}$ are determined to be:
\ben
&& B^{(1)}_{y_2}=\frac{\p_{y_3}\psi_2}{\psi_3}\,,\qquad B^{(1)}_{y_3}=-\frac{\p_{y_2}\psi_3}{\psi_2} \, , \\
&&B^{(2)}_{y_1}=-\frac{\p_{y_3}\psi_1}{\psi_3}\,,\qquad B^{(2)}_{y_3}=\frac{\p_{y_1}\psi_3}{\psi_1} \, , \\&&
B^{(3)}_{y_1}=\frac{\p_{y_2}\psi_1}{\psi_2}\,,\qquad B^{(3)}_{y_2}=-\frac{\p_{y_1}\psi_2}{\psi_1} \, . \label{B3mu}
\een
Thus all components of $B^{(a)}_\mu$ are determined in terms of derivatives of $\psi_a$ 
except~$B^{(a)}_{\mu} \Phi^{(a)\mu}$.

From \eqref{VEqu1}, we get following integrability condition. With~$D_i\Theta=\p_i\Theta+A_i$, 
\be
\p_t D_i\Theta=0\, , \qquad \p_i \big(\frac{\sqrt{\tilde g}}{R^2}D^i\Theta \big)=0\,,  \qquad i=1,2,3 \, .
\ee
Here $\tilde g$ is the determinant of three dimensional metric and indices are raised and lowered by three 
dimensional metric. The above condition together with \eqref{XEqu2} immediately implies that
\be
R(A)_{ti}=0 \, , 
\ee
that is, $R(A)_{\mu\nu}$ has only magnetic components.

As in \S\ref{Weyl}, we must also impose the dilatation gauge condition, namely that we should fix 
the determinant to its asymptotic value. 
The condition $R^{2}=\rho^{2}$ (see equation~\eqref{Reqrho}) is now replaced by
\be
R^{2} =  \psi_{1} \psi_{2} \psi_{3} \, \rho^{2} \,.
\ee

Finally, we have another set of BPS equations coming from variation of the auxiliary fermions~$\chi^i$:
\be
\delta\chi^i=-\frac{1}{12}\Gamma_a\Gamma_b\Gamma^\mu D_\mu T^{abij}\epsilon_j+\frac{1}{6} R(V)^i{}_{j\mu\nu}\Gamma^\mu\Gamma^\nu\epsilon^j 
-\frac{i}{3} R(A)_{\mu\nu}\Gamma^\mu\Gamma^\nu\epsilon^i+D \epsilon^i \, .
\ee
After some tedious work, one can show that these BPS equations do not give any new 
constraints on $R(V)^i{}_j$ and $R(A)_{\mu\nu}$ and only determines scalar field $D$ in terms 
of $T_{\mu\nu}$, $R(V)^i{}_j$ and $R(A)_{\mu\nu}$. 

This finishes the general analysis of the gravity multiplet, we see that the gauge fields are partially 
but not completely constrained by the supersymmetry analysis. We will next couple the gravity multiplet 
to vector multiplets. We see immediately that the~$SU(2)$ gauge field is not coupled to the vector multiplets,
and therefore there cannot be any more constraints on it. 
In particular, we see that the most general solution for the $SU(2)$ gauge field is given in terms of 
the $SU(2)$ triplet of one forms $\Phi^{(a)}$ in equations~\eqref{Bphitcomp}--\eqref{B3mu}.

The $U(1)$ gauge field on the other hand, couples to the vector multiplet fields, and we turn to this 
analysis next. Since we have completely analyzed the~$SU(2)$ gauge field above, we shall 
now set it to zero to investigate if the~$U(1)$ gauge field is localized further.
In this case, the triplet of one forms are exact as before. 
In the vector multiplet analysis, there are two possible sources of changes:
\vspace{-0.2cm}
\begin{enumerate}
\item Partial derivatives $\p_{\mu}$ are replaced by covariant derivatives $D_{\mu}$. 
Note here that the Euclidean continuation defines 
$$D_{\mu} X^{I} = \p_{\mu} X^{I} + A_{\mu} X^{I} \, , \qquad 
D_{\mu} \bar{X^{I}} = \p_{\mu} \bar{X^{I}} - A_{\mu} \bar{X^{I}} \, , $$
so that we can do the above replacement even after the decomposition into real and imaginary parts. 
\vspace{-0.2cm}
\item The values of the various bilinears $a^{\mu}$, $b^{\mu}$ etc may change. 
\end{enumerate}

We now discuss the various bilinears. The bilinears can change since the 
field $\Theta$ as well as the gauge field $A_{\mu}$ are no longer zero. 
Turning on the field $\Theta$ is equivalent to a $U(1)$ rotation, and so all the charged 
quantities of the gravity multiplet rotate accordingly, including the Killing spinors. 
Since in the vector multiplet equations, the spinors are not differentiated, the 
equations will all rotate by the same $\Theta$ dependent factor, and there is 
no net effect of $\Theta$ on the equations. 

On turning on $A_{\mu}$, 
the fields $a^{\mu}$, $b_{\mu}$, and $\wh b_{\mu}$ \eqref{defamu} are the Euclidean 
counterparts of the Killing vector $K^{\mu}$ and the one-form $\Phi^{(\a)}_{\mu}$, 
and therefore do not change. 
The metric, however, does change in that $g_{0\mu}$ components now contain the 
vector field~$V_{\mu}$. In the BPS equations in~\S\ref{vecdetails}, the fields~$a^{\mu}$,~$b_{\mu}$, 
and~$\wh b_{\mu}$ are always contracted with the covariant derivative~$D_{\mu}$, and therefore the only 
possible $V_{\mu}$ dependence appears in a combination with~$D_{0}$ which, in fact, 
kills the scalar fields. Therefore, there is no~$V_{\mu}$ dependence coming through the 
vectors and one-forms.
There are also tensor fields appearing in the BPS equations, which can depend on~$V_{\mu}$.
However, using the fact that the tensor fields $M_{ab}$, $N_{ab}$ etc are determined algebraically 
in terms of the vectors and one-forms, we can show that the tensors with tangent space 
indices e.g.~$M_{ab}$ do not depend on~$V_{\mu}$. We show this in Appendix~\ref{Mabanal}.

The form of the equations (\ref{1stBPSequ1}--\ref{2ndBPSequ4})
(and therefore  equations (\ref{fSol.1}--\ref{HSol.2})) 
retain the same structure with the same numerical values for the various bilinears 
(\ref{defamu}--\ref{Npr01}), but there are explicit changes to the the various quantities 
entering these equations. In particular, the field $f_{ab}$ is replaced by a combination of $f_{ab}$
and the gauge field, while the scalars $H^{I}$, $J^{I}$ and $K^{I}$ enter as before. 

We will now solve this modified system of equations. 
Firstly, note that from the definition of the covariant derivative in Euclidean space,
we have 
$$D_{\mu} H = \p_{\mu} H + A_{\mu} J \, , \qquad D_{\mu} J = \p_{\mu} J + A_{\mu} H \, .$$
We notice that the same combination of the gauge fields $A_{\mu}$ appears in 
\eqref{JSol.1} and \eqref{HSol.2}. We set this combination to zero by a gauge choice:
\be\label{gaugechoice}
\cosh\eta\sin\psi \, A_\eta +\sinh\eta\cos\psi\, A_\psi  = 0 \, .
\ee 
We can now solve \eqref{JSol.1} and \eqref{HSol.2} as before to get:
\be
\label{HJnewsol}
H=\frac{C(\theta,\phi)}{\cosh\eta}  \, , \qquad J=0 \, .
\ee
Note that $C$ is a function of both $\theta$ and $\phi$ so far.
Replacing the derivatives by covariant derivatives in \eqref{Sol.1}, we get
\be\label{Sol.1new}
(D_{\theta} -D_{\phi})  H = 0 \, , \qquad  (D_{\theta} -D_{\phi}) \, J = 0 \, . 
\ee
Plugging in $J=0$ in these equations gives
\be \label{AthAphi}
A_{\theta}-A_{\phi} =0 \, , \qquad \p_{\theta-\phi} C(\theta,\phi) =0 \, ,
\ee
which means that we can write $C(\theta+\phi)$ as before. 

The equations \eqref{pthphiH}, \eqref{pthphiJ} change to 
\bea\label{pthphiHnew}
D_{\theta+\phi} H + \frac{\sin\psi\sinh\eta}{\sin^2\psi+\sinh^2\eta} \frac{K_1-K_2}{2}\cos\psi= 0 \, , \\
\label{pthphiJnew} 
D_{\theta+\phi} J + \frac{\sin\psi\sinh\eta}{\sin^2\psi+\sinh^2\eta} \frac{K_1-K_2}{2}\cosh\eta= 0 \, . 
\eea
Plugging in the values of $H$ and $J$ into the first equation, and multiplying by $\cosh \eta$,
we find two terms in the first equation which are both  independent of $\eta$, 
and therefore should take their values at $\eta \to \infty$. We find:  
\be
K_{1}-K_{2}=0 \, , \qquad \p_{\theta+\phi} C(\theta,\phi) =0 \, ,
\ee
which implies that $C$ is a constant. 
Plugging in the values of $H$ and $J$ in the second equation then gives us:
\be \label{AthAphipl}
A_{\theta}+A_{\phi} =0  \, .
\ee

The equations \eqref{JSol.2}, \eqref{KSol} change to 
\bea\label{JSol.2new}
-2\sinh\eta\cos\psi D_\eta J+2\sin\psi\cosh\eta D_\psi J-2H+2J\cosh\eta\cos\psi+K_{3} \cosh\eta = 0  \, , \\
\label{KSolnew}  
-2\sinh\eta\cos\psi D_\eta H+2\sin\psi\cosh\eta D_\psi H-2H\cosh\eta\cos\psi+2J+K_{3}  \cos\psi = 0 \, .
\eea
Plugging in the values of $H$ and $J$ in \eqref{KSolnew} 
gives $K_{3}$ as a function of $H$ exactly as before, and plugging that into~\eqref{JSol.2new} 
gives us
\be
-2\sinh\eta\cos\psi A_\eta +2\sin\psi\cosh\eta A_\psi  =0\, . 
\ee
Combining this with \eqref{gaugechoice}, \eqref{AthAphi}, and \eqref{AthAphipl}, we get:
\be
A_{\mu} =0 \, , \qquad (\mu = \eta, \theta, \psi, \phi ) \, .
\ee
Thus we see that the~$U(1)$ gauge field is constrained to vanish upon coupling to 
vector multiplets. As mentioned in the introduction, it is possible that the~$SU(2)$
gauge field also gets constrained upon coupling to hypermultiplets. However, we note that,
in both these cases, the points on the localization manifold where the matter multiplets vanish are singular, and
new branches of solutions parameterized by the auxiliary gauge fields open up. 
These subtleties need to be taken into account while performing the functional integral.

\section*{Acknowledgements}

It is a pleasure to acknowledge useful discussions with Atish Dabholkar, Jo\~ao Gomes, 
Bindusar Sahoo, Ashoke Sen and Bernard de Wit. 
We thank Satoshi Nawata and Ashoke Sen for comments on a draft of the paper. 
This work is supported by the ERC Advanced Grant no. 246974,
{\it ``Supersymmetry: a window to non-perturbative physics''}.

\appendix

\section{Smoothness and the one-form $V$ \label{smoothV}}

In this appendix, we show that the one form $V$ appearing in the metric \eqref{genmet} 
can be set to zero using the smoothness criterion. 
We work in the gauge \eqref{detgcond} which gives us:
\be
R(y^{\a}) = \rho \, . 
\ee 
The integrability condition \eqref{integrabilityV} becomes:
\be
\rho^2 \, \p_\rho^2 \, \Theta+\Box \, \Theta  \=  0 \, , 
\ee
where $\Box$ is the Laplacian on the unit 2-sphere.
Expanding in $S^{2}$ spherical harmonics,
\be
\Theta(\rho,\psi,\phi) \= \sum_{\ell,m}\Theta_{\ell m}(\rho) \, Y_{\ell m} (\psi,\phi) \, , \qquad 
\Box \, Y_{\ell m} = -\ell (\ell+1)\, Y_{\ell m} \, , 
\ee
we get the equation 
\be
\rho^2 \, \frac{d^{2}}{d\rho^{2}} \, \Theta_{\ell m}-\ell (\ell+1) \, \Theta_{\ell m}=0 \, ,
\ee
which has a two-dimensional space of solutions
\be\label{cldl}
\Theta_{\ell m}(\rho) =\frac{c_\ell}{\rho^\ell}+d_\ell \, \rho^{\ell+1} \, . 
\ee

Now we do an analytic continuation to Euclidean space as in the main text \eqref{eads2s2}.
Imposing the boundary condition that the metric \eqref{genmet} 
equals the metric \eqref{eads2s2} as $\eta \to \infty$ kills all the $c_{\ell}$s 
and $d_{\ell}$s except $c_{0}$. An easy way to see this is to compute the Ricci scalar curvature of the 
metric \eqref{genmet}, which is invariant under coordinate transformations. The points $\rho=0$
and $\rho=\infty$ are mapped to the points $w = \pm 1$, which is on the boundary of the 
unit disk ($\eta = \infty$), which 
should have vanishing Ricci scalar curvature by our boundary conditions above. 
The Ricci scalar curvature of the modes $c_{\ell}, \ell >0$ blows up for $\rho \to 0$,
and the Ricci scalar curvature of the modes $d_{\ell}, \ell >0$ blows up for $\rho \to \infty$.
This implies that the only mode which is allowed in \eqref{cldl} is the constant mode $\Theta_{00}$. 
Plugging this result into \eqref{dVeqn}, we deduce that $V$ is a locally exact form, 
and can therefore be absorbed in the definition of the time coordinate~$t$.

\section{Algebraic relations between the various bosonic fields \label{Mabanal}}

In this section we compute the relation between the tensor field~$M_{\mu\nu}$ and the Killing vector
and one-forms. We show that when we turn on R-symmetry $U(1)$ gauge field, the components of the 
tensor bilinear $M^{\mu\nu}$ have a dependence on the field~$V_\mu$ appearing in the metric. However, 
the $M^{ab}$ with flat indices do not have a~$V_\mu$ dependence. 
A similar conclusion holds for other tensor bilinears, we suppress the details since the analysis is very similar. 
We have
\be
M^{\mu\nu}=\psi^\dagger\Gamma^{\mu\nu}\psi \, . 
\ee
The killing vector and exact form are
\be
K^\mu=\psi^\dagger\Gamma^{\mu}\psi,\qquad b_\mu=\psi^\dagger\Gamma_{\mu}\Gamma_5\psi \, .
\ee
Then by using Fierz identities, we get
\ben
K_\mu M^{\mu\nu}=f_2b^\nu \, , \qquad K_\mu \tilde M^{\mu\nu}=b^\nu f_1 \, ,
\een
with 
\be
f_1=\psi^\dagger\psi,\qquad f_2=\psi^\dagger\Gamma_5\psi \, . 
\ee
Thus we can write
\be
M^{\mu\nu}=\frac{1}{K^2}[f_2(K^\mu b^\nu-K^\nu b^\mu)+f_1K_\alpha b_\beta\varepsilon^{\alpha\beta\mu\nu}] \, .
\ee

For our metric
\be
ds^2=-R^2(dt+V)^2+\frac{1}{R^2}dy^\alpha dy^\alpha \, , 
\ee
the vielbeins are
\be
e^0=R(dt+V),\quad e^{\hat a}=\frac{1}{R}dy^{\hat a},\quad \hat a=1,2,3 \, .
\ee
One can easily compute the components~$M^{ab}$ to be:
\be
M^{0\hat a}=-\frac{f_2}{R^2}b^ie^{\hat a}_i,\qquad M^{\hat a\hat b}=-f_1R^2\varepsilon^{ijk}b_ie^{\hat a}_je^{\hat b}_k \, ,
\ee
and we see that, as claimed above, they are indeed independent of the field~$V_{\mu}$.

\section{Details of the vector multiplet solutions \label{vecdetails}}

As in the main text, we will suppress the index $I$ in the following.  
We begin with the chiral equation \eqref{1stBPSequ}. In order to produce equations 
for the fermion bilinears, we multiply it by various spinors on the left. Using the 
spinors $(\xi^+_{++})^\dagger$, $(\xi^+_{++})^\dagger\Gamma_5$, $(\xi^-_{--})^\dagger$,
and $(\xi^-_{--})^\dagger\Gamma_5$, we get the four equations:
\be\label{1stBPSequ1}
\frac{1}{2}f_{ab}M^{ab}+2ia^\mu\p_\mu H-2ib^\mu\p_\mu J-8H+8J\cosh\eta\cos\psi+4K_{3}\cosh\eta=0  \, , \\ 
\ee
\be\label{1stBPSequ2}
\frac{1}{2}f_{ab}\widetilde M^{ab}-2ib^\mu\p_\mu H+2ia^\mu\p_\mu J-8H\cosh\eta\cos\psi+8J+4K_{3}\cos\psi=0 \, , \\ 
\ee
\be\label{1stBPSequ3}
\frac{1}{2}f_{ab}N^{ab}-2i\widehat b^\mu\p_\mu J-8e^{i(\theta+\phi)} J\sinh\eta\sin\psi +4iK_1e^{i\beta}\cosh\eta =0 \, , \\ 
\ee
\be\label{1stBPSequ4}
\frac{1}{2}f_{ab}\widetilde N^{ab}-2i\widehat b^\mu\p_\mu H+8e^{i(\theta+\phi)} H\sinh\eta\sin\psi +4iK_1e^{i\beta}\cos\psi=0 \, . 
\ee

Similarly, we obtain a set of BPS equations from \eqref{2ndBPSequ}. Multiplying from the 
left with spinors  $(\xi^-_{--})^\dagger$, $(\xi^-_{--})^\dagger\Gamma_5$, $(\xi^+_{++})^\dagger$,
and $(\xi^+_{++})^\dagger\Gamma_5$, we get the following four equations:
\be\label{2ndBPSequ1}
\frac{1}{2}f_{ab} {M'}^{ab}+2i {a'}^\mu\p_\mu H-2i{b'}^\mu\p_\mu J+8H-8J\cosh\eta\cos\psi-4K_{3}\cosh\eta=0  \, , \\ 
\ee
\be\label{2ndBPSequ2}
\frac{1}{2}f_{ab}\widetilde M'^{ab}-2ib'^\mu\p_\mu H+2ia'^\mu\p_\mu J+8H\cosh\eta\cos\psi-8J-4K_{3}\cos\psi=0 \, , \\ 
\ee
\be\label{2ndBPSequ3}
\frac{1}{2}f_{ab}N'^{ab}-2i\widehat b'^\mu\p_\mu J-8 e^{-i(\theta+\phi)} J\sinh\eta\sin\psi +4iK_2e^{i\alpha}\cosh\eta=0 \, , \\ 
\ee
\be\label{2ndBPSequ4}
\frac{1}{2}f_{ab}\widetilde N'^{ab}-2i\widehat b'^\mu\p_\mu H+8 e^{-i(\theta+\phi)} H\sinh\eta\sin\psi +4iK_2e^{i\alpha}\cos\psi=0 \, . 
\ee   

Here, we have used various Euclidean fermion bilinears 
whose explicit values 
for our choice of Killing spinors are presented below.

\ndt The various vector fields are
\bea \label{defamu}
a_\mu=(\xi^+_{++})^\dagger \Gamma_\mu\xi^+_{++} \, & : &  \qquad  
a_\theta=4\sinh^2\eta \,,\quad a_\eta=0 \, , \quad a_\phi=-4\sin^2\psi\,,\quad a_\psi=0 \, . \\
b_\mu=(\xi^+_{++})^\dagger \Gamma_\mu\Gamma_5\xi^+_{++} \, & : &  \qquad b_\theta=0\,,\quad 
b_\eta=-4i\sinh\eta \cos\psi\,,\quad b_\phi=0\,,\quad b_\psi=4i\sin\psi\cosh\eta \, . \nonumber 
\eea
\ben
\widehat b_\mu=(\xi^-_{--})^\dagger \Gamma_\mu\Gamma_5\xi^+_{++} \, : && 
\widehat b_\theta=-4e^{i(\phi+\theta)}\sinh\eta\sin\psi\,,\quad \widehat b_\eta=4ie^{i(\phi+\theta)}\cosh\eta\sin\psi \, , \nonumber \\
&& \widehat b_\phi=-4e^{i(\phi+\theta)}\sin\psi\sinh\eta \,, \quad \widehat 
b_\psi=4ie^{i(\phi+\theta)}\sinh\eta\cos\psi \, . \qquad 
\een
\bea
\widehat b'_\mu=(\xi^+_{++})^\dagger \Gamma_\mu\Gamma_5\xi^-_{--} \, : 
&&\widehat b'_\theta=4e^{-i(\phi+\theta)}\sinh\eta\sin\psi,\quad \widehat b'_\eta=4ie^{-i(\phi+\theta)}\cosh\eta\sin\psi\, , \cr
&& \widehat b'_\phi=4e^{-i(\phi+\theta)}\sin\psi\sinh\eta\,, \quad \widehat b'_\psi=4ie^{-i(\phi+\theta)}\sinh\eta\cos\psi \, . \qquad 
\eea
\be
a'_\mu=(\xi^-_{--})^\dagger \Gamma_\mu\xi^-_{--}=a_\mu \, , \qquad 
b'_\mu=(\xi^-_{--})^\dagger \Gamma_\mu\Gamma_5\xi^-_{--}=-b_\mu \, .
\ee
\be
(\xi^+_{++})^\dagger \G^{\mu} \xi^-_{--}  = 0 \, .
\ee

\ndt The various scalars fields are
\bea
&& (\xi^+_{++})^\dagger \xi^+_{++} = (\xi^-_{--})^\dagger \xi^-_{--} = 4 \cosh \eta  \, , \qquad (\xi^+_{++})^\dagger \G^{5} \xi^+_{++} = (\xi^-_{--})^\dagger \G^{5} \xi^-_{--} = 4 \cos \psi  \, , \cr
&& (\xi^+_{++})^\dagger \xi^-_{--}  = (\xi^+_{++})^\dagger \G^{5} \xi^-_{--} = 0 \, .
\eea

\ndt The various tensor fields are:
\ben
M^{ab}=(\xi^+_{++})^\dagger \Gamma^a\Gamma^b\xi^+_{++} \, : 
&& M^{01}=-4i,\quad M^{02}=0\,,\quad M^{03}=0 \,, \quad M^{12}=4i\sinh\eta\sin\psi\ \, , \cr
&& M^{13}=0 \, , \quad M^{23}=-4i\cosh\eta\cos\psi \, .
\een
\ben
\widetilde M^{ab}=(\xi^+_{++})^\dagger \Gamma_5\Gamma^a\Gamma^b\xi^+_{++} \, : 
&& \widetilde M^{01}=-4i\cosh\eta\cos\psi\,,\quad \widetilde M^{02}=0\,,\quad \widetilde 
M^{03}=4i\sinh\eta\sin\psi \, ,\nonumber\\
&& \widetilde M^{12}=0\,,\quad \widetilde M^{13}=0\,,\quad \widetilde M^{23}=-4i \, . 
\een
\be
M'^{ab}=(\xi^-_{--})^\dagger \Gamma^a\Gamma^b\xi^-_{--}=-M^{ab} \, , \qquad 
\widetilde M'^{ab}=(\xi^-_{--})^\dagger \Gamma_5\Gamma^a\Gamma^b\xi^-_{--}=-\widetilde M^{ab} \, .\\
\ee

\ben
N^{ab}=(\xi^-_{--})^\dagger \Gamma^a\Gamma^b\xi^+_{++} \, :
&& N^{01}=0,\quad N^{02}=-4e^{i(\phi+\theta)}\cos\psi,\quad N^{03}=4ie^{i(\phi+\theta)}\, , \nonumber \\
&& N^{12}=4ie^{i(\phi+\theta)}\cosh\eta\cos\psi\,,\quad N^{13}=4e^{i(\phi+\theta)}\cosh\eta\,, \qquad  \nonumber \\
&& N^{23}=4ie^{i(\phi+\theta)}\sinh\eta\sin\psi \, . 
\een
\ben
\widetilde N^{ab}=(\xi^-_{--})^\dagger \Gamma_5\Gamma^a\Gamma^b\xi^+_{++}\, : 
&& \widetilde N^{23}=0\,,\quad \widetilde N^{13}=4e^{i(\phi+\theta)}\cos\psi,\quad \widetilde N^{12}=4ie^{i(\phi+\theta)}\cr
&& \widetilde N^{03}=4ie^{i(\phi+\theta)}\cosh\eta\cos\psi\,,\quad \widetilde N^{02}=-4e^{i(\phi+\theta)}\cosh\eta\,, \nonumber\qquad \\
&& \widetilde N^{01}=4ie^{i(\phi+\theta)}\sinh\eta\sin\psi \, .
\een
\ben
N'^{ab}=(\xi^+_{++})^\dagger \Gamma^a\Gamma^b\xi^-_{--}\, : 
&& N'^{01}=0\,,\quad N'^{02}=4e^{-i(\phi+\theta)}\cos\psi,\quad N'^{03}=4ie^{-i(\phi+\theta)} \,,\nonumber \\
&& N'^{12}=4ie^{-i(\phi+\theta)}\cosh\eta\cos\psi\,,\quad N'^{13}=-4e^{-i(\phi+\theta)}\cosh\eta\,, \nonumber\\
&& N'^{23}=4ie^{-i(\phi+\theta)}\sinh\eta\sin\psi \, . 
\een
\ben 
\widetilde N'^{ab}=(\xi^+_{++})^\dagger \Gamma_5\Gamma^a\Gamma^b\xi^-_{--}\, : 
&& \widetilde N'^{23}=0\,,\quad \widetilde N'^{13}=-4e^{-i(\phi+\theta)}\cos\psi \,,\quad \widetilde N'^{12}=4ie^{-i(\phi+\theta)} \, , \nonumber \\
&& \widetilde N'^{03}=4ie^{-i(\phi+\theta)}\cosh\eta\cos\psi,\quad \widetilde N'^{02}=4e^{-i(\phi+\theta)}\cosh\eta \,, \nonumber \\
&& \widetilde N'^{01}=4ie^{-i(\phi+\theta)}\sinh\eta\sin\psi \, . \label{Npr01}
\een

\providecommand{\href}[2]{#2}\begingroup\raggedright\endgroup

\end{document}